\documentstyle[12pt,epsf,psfig]{article}
\begin{document}
\baselineskip 0.6cm
\newcommand{\gsim}{ \mathop{}_{\textstyle \sim}^{\textstyle >} }
\newcommand{\lsim}{ \mathop{}_{\textstyle \sim}^{\textstyle <} }
\newcommand{\vev}[1]{ \langle {#1} \rangle }
\newcommand{\EV}{ {\rm eV} }
\newcommand{\KEV}{ {\rm ~keV} }
\newcommand{\MEV}{ {\rm ~MeV} }
\newcommand{\GEV}{ {\rm ~GeV} }
\newcommand{\TEV}{ {\rm ~TeV} }
\newcommand{\eps}{\varepsilon}
\newcommand{\barr}[1]{ \overline{{#1}} }
\newcommand{\del}{\partial}
\newcommand{\nn}{\nonumber}
\newcommand{\ra}{\rightarrow}
\newcommand{\bino}{\tilde{\chi}}
\def\tr{\mathop{\rm tr}\nolimits}
\def\Tr{\mathop{\rm Tr}\nolimits}
\def\Re{\mathop{\rm Re}\nolimits}
\def\Im{\mathop{\rm Im}\nolimits}
\setcounter{footnote}{0}

\begin{titlepage}

\begin{flushright}
CERN-TH/2002-172\\
UT-02-43
\end{flushright}

\vskip 3cm
\begin{center}
{\large \bf Baryogenesis and Gravitino Dark Matter\\
in Gauge-Mediated\\
Supersymmetry-Breaking Models}
\vskip 2.4cm

\center{Masaaki~Fujii$^{a,b}$ and T.~Yanagida$^{a,b,}$}\footnote{On
leave from University of Tokyo.}\\
$^{a}${\it{CERN Theory Division, CH-1211 Geneva 23, Switzerland }}\\
$^{b}${\it{Department of Physics, University of Tokyo, Tokyo 113-0033,
Japan}}\\
\vskip 2cm

\abstract{We discuss two cosmological issues in a generic
  gauge-mediated supersymmetry (SUSY)-breaking model, namely the
  Universe's baryon asymmetry and the gravitino dark-matter density.  We
  show that both problems can be simultaneously solved if there
  exist extra matter multiplets of a SUSY-invariant mass of the order
  of the ``$\mu$-term'', as suggested in several realistic SUSY 
  grand-unified theories. We propose an attractive scenario in which the
  observed baryon asymmetry is produced in a way 
  totally independent of the
  reheating temperature of inflation without causing any 
  cosmological gravitino problem.  Furthermore, in a relatively wide
  parameter space, we can also explain the present mass density of
  cold dark matter by the thermal relics of the gravitinos without an
  adjustment of the reheating temperature of inflation.  We  point
  out that there is an interesting relation between the baryon
  asymmetry and the dark-matter density.}

\end{center}
\end{titlepage}

\renewcommand{\thefootnote}{\arabic{footnote}}
\setcounter{footnote}{0}
%
%
%
\section{Introduction}
The mediation of supersymmetry (SUSY)-breaking effects
to the standard-model (SM) sector is a fundamental ingredient in
the SUSY extension of the SM.
Among various models proposed so far, the gauge mediation model
with dynamical SUSY breaking (GMSB)~\cite{GMSB} is the most attractive.
It may not only solve the flavour-changing neutral-current
(FCNC) problem, but also provide a natural explanation of the
large hierarchy between the electroweak scale and the Planck scale
because of the dynamical nature of SUSY breaking.
Extensive analyses of the GMSB models have been performed on the physics
at the  ${\mbox {TeV}}$ scale.

The remaining important task is to construct
a successful cosmology in the GMSB models.
The biggest difficulty comes from a stringent upper bound on
the reheating temperature of inflation $T_{R}$
in order to avoid the overproduction of gravitinos~\cite{G-problem}.
The upper bound on $T_{R}$ is about $10^{6}\GEV$ when $m_{3/2}\simeq
10\MEV$ for instance, and it even reaches $T_{R}\lsim 10^{3}\GEV$ 
in the case of
the light gravitino $m_{3/2}\lsim 100\KEV$.
In addition, we have to adjust the reheating temperature 
to just below this  upper bound if we want to explain the 
required mass density of cold dark matter (CDM) by
the gravitino LSPs (lightest SUSY particles).
Furthermore,
we also 
have to explain the required baryon asymmetry with just the same
reheating temperature.
These facts severely constrain models of inflation and 
baryogenesis mechanism, and require a doubtful conspiracy
between these two independent  processes.

Even only explaining the required baryon asymmetry is very difficult.
The unique mechanism to generate the baryon asymmetry in such low
reheating temperatures, $T_{R}\lsim$ $10^6\GEV$, has been considered
to be the Affleck--Dine (AD) baryogenesis~\cite{AD}.~\footnote{In the
  case of a heavier gravitino $m_{3/2}\gsim 10\MEV$, the leptogenesis
  through inflaton decays~\cite{inflaton-decay} or a coherently
  oscillating right-handed sneutrino~\cite{sneutrino-dominance} can
  explain the observed baryon asymmetry.}  However, in recent
developments, it becomes clear that the coherent oscillation of the
flat direction field (AD field) is unstable against spatial
perturbations and that this field fragments into non-topological
solitons called Q-balls~\cite{Coleman} after dozens of its
oscillations.  In GMSB models, it is known that the resultant Q-balls
are absolutely stable, and this fact has motivated some works to
relate the observed baryon asymmetry and the mass density of dark
matter~\cite{Q-ball-org}.  Now that some detailed lattice
simulations~\cite{KK} have been performed, we can easily calculate the
typical size and the number density of the Q-balls generated in GMSB
models. Unfortunately, these simulations have revealed that the
produced Q-balls generally overclose the Universe, which means that
the usual AD baryogenesis is not a phenomenologically viable scenario
as the origin of the observed baryon asymmetry in the GMSB models.

A possible solution to the difficulty is to adopt the AD
field, which has a SUSY-invariant mass term.
This is  the case for the AD
leptogenesis via the $LH_{u}$ flat direction~\cite{MY-L}. The scenario can explain the
baryon asymmetry provided that the ``$\mu$-term'' exists even in the
very high energy
(above the intermediate) scale, and that the
mass of the lightest neutrino is extremely small~\cite{AFHY}.~\footnote{This fact
allows us to have definite predictions on the rate of neutrinoless
double beta $(0\nu\beta\beta)$ decay, which will be tested in
future experiments~\cite{double-beta}.}
However, we need a conspiracy again between the model of inflation and
the leptogenesis if we want to 
explain the gravitino dark matter, simultaneously.

In this paper, we propose a solution to both problems.
We consider the cosmology in GMSB models under
the existence of extra matter multiplets of a SUSY-invariant mass of the
order of the $\mu$-term. The existence of extra matter multiplets,
such as those transforming $\bf{5}+\bf{\bar{5}}$ under the SU(5)$_{\rm
GUT}$ group, is indicated by
several realistic SUSY GUT models~\cite{SUSY-GUT} 
in which we can naturally solve the
doublet--triplet splitting problem by utilizing some discrete symmetries.

As suggested from the  AD leptogenesis, adopting
an extra matter field as a member of the AD fields
opens up various possibilities to construct successful AD baryogenesis scenarios.
In the present work, we will propose the most attractive one,
which can generate the required baryon asymmetry in a way totally independent
of the reheating temperature of inflation,
without any conflicts with the constraint 
on cosmological gravitino abundance.
Surprisingly, we can explain simultaneously, 
in a relatively wide parameter space, the required
mass density of CDM by the thermal relics of the gravitino LSPs
without an adjustment of the reheating temperature of inflation.
In this region, we derive a novel relation between the mass density
of the baryon and that of dark matter,
which is written solely by low-energy parameters.
We will also discuss the non-thermal gravitino production via  late-time
decays of the {\it unstable} Q-balls, and its phenomenological
implications and experimental constraints.

\section{The Model}
In the present scenario, we assume the existence of a pair of extra
matter multiplets, ${\bf{5}}_{x}$ and ${\bf{\bar{5}}}_{x}$, which have
a SUSY-invariant mass term in the superpotential, $W\supset
m_{x}{\bf{5}}_{x}{\bf{\bar{5}}}_{x}$.~\footnote{The following
  arguments can also be applied to the case where the extra matter
  fields belong to ${\bf 10}+{\bf \barr{10}}$ representations of the
  SU(5)$_{\rm GUT}$.}  In the following discussion, we assume that the
size of $m_{x}$ is of the order of the $\mu$-term $(|\mu|\sim
1\TEV)$, as suggested in realistic SUSY GUT models~\cite{SUSY-GUT}.

By adopting one of these extra fields as an AD field, this mass term
supplies a quadratic potential along the relevant flat direction, which
forbids the formation of stable Q-balls in the GMSB models. This is not the case, however,
when the mass term is generated by an expectation value of some field
with a renormalizable coupling to the extra matter multiplets, such as
$W\supset S{\bf{5}}_{x}{\bf{\bar{5}}}_{x}$, where the $S$ field is to
obtain an expectation value. If this is the case, the $S$
field is driven to the origin when the AD field develops a large
amplitude, and then the required quadratic potential vanishes.

In order to avoid the formation of stable Q-balls in the
GMSB models, the mass term should be present even after
the AD field obtains a large amplitude. This condition can be easily
satisfied, for example, if the mass term is generated by some
non-renormalizable operator~\cite{TY-mu}, $W\supset
(S^{2}/M_{*}){\bf{5}}_{x}{\bf{\bar{5}}}_{x}$,
where $M_{*}\equiv 2.4\times 10^{18}\GEV$ is the reduced
Planck scale. We assume that  this is the case in the remainder of this paper.

Our model of the AD baryogenesis is basically the same as the one
proposed in Section II B of Ref.~\cite{FH}. We adopt the flat direction
labelled by a linear combination of the monomials of the chiral
superfields: $\bar{U}\bar{D}\bar{D}_{x},\;Q\bar{D}_{x}L$.  Here, $Q$,
$\bar{U}$, $\bar{D}$ and $L$ denote superfields (and their scalar
components) of left-handed quark doublet, right-handed up- and
down-type quarks and left-handed lepton doublet, respectively;
$\bar{D}_{x}$ belongs to the ${\bf{\bar{5}}}_{x}$ multiplet and has
the same gauge quantum numbers as $\bar{D}$.  As for an explicit
parametrization of the flat direction, see Ref.~\cite{FY}.

The relevant baryon-number (B)-violating operators are supplied in the
K\"ahler potential. (The effects of possible non-renormalizable 
operators in the superpotential will be discussed later.)
We take the following operators, for example, which
are consistent with the $R$-symmetry:
\begin{equation}
 \delta K=\lambda_{1}\frac{1}{M_{*}^{2}}Q\bar{U}^{\dag}\bar{D}^{\dag}_{x}L+
\lambda_{2}\frac{1}{M_{*}^{2}}Q\bar{U}^{\dag}\bar{D}^{\dag}L+{\rm h.c.}\;,
\label{kahler}
\end{equation}
where $\lambda_{i}$'s are coupling constants.
The SUSY-breaking effect due to the non-zero energy density of the
Universe induces the following term in the scalar potential, which
provides the AD field with a phase rotational motion:
\begin{equation}
 \delta V= 3H^2\;\delta K\;,
\label{sugra-effect}
\end{equation}
where $H$ is the Hubble parameter of the expanding Universe.  Here,
the fields in $\delta K$ represent the scalar components of
corresponding superfields.~\footnote{ The terms induced by the
  SUSY-breaking effect in the true vacuum have only a negligible role in
  the present scenario. Note that the existence of $\delta V$ in
  Eq.~(\ref{sugra-effect}) is independent of the quantum charge of the
  inflaton.  This is in clear contrast with the model proposed in
  Ref.~\cite{Mazumdar}, which requires the inflaton $I$ to be a singlet in
  order to write relevant interaction terms, $\delta
  K=(I/M_{*})\phi^{\dag}\phi$ and $\delta W=(I/M_{*}){\cal
    W}_{\alpha}{\cal W}^{\alpha}$ to solve the Q-ball problem in the
  GMSB models. Such interactions generally lead to the high reheating
  temperature of inflation, which results in the overproduction of the
  gravitinos.  In addition, there is no definite 
description of the available flat direction in the MSSM
and the prediction of the resultant baryon asymmetry. It is not clear to us 
whether their scenario can actually produce the observed baryon asymmetry
without a conflict with the cosmological gravitino problem.}

The total scalar potential of the AD field $\phi$ can be written as follows:
\begin{equation}
 V=(m_{x}^2-c_{H}H^2)|\phi|^2+\frac{H^2}{4 M_{*}^2}(a_{H}\phi^4+{\rm h.c.})+\ldots\;,
\label{scalar-potential}
\end{equation}
where the ellipsis denotes the higher-order terms coming from the
K\"ahler potential;
$a_{H}$ is a complex coupling constant and
$c_{H}={\cal{O}}(1)$ is a real positive coefficient. The definition of
$\phi$ is the same as that in Ref.~\cite{FY}, which denotes the complex
scalar field along the flat direction. The second term is induced from
the four-point couplings between the inflaton $I$ and the AD
field in the K\"ahler potential 
($K\supset I^{\dag}I\phi^{\dag}\phi/M_{*}^2$)~\cite{DRT}.

\section{Baryon Asymmetry}
\label{B-asymmetry}

The evolution of the AD field and the resultant baryon asymmetry are
almost the same as those presented in Refs.~\cite{FH,FY}.
Here, we briefly review the important points. The AD field
is assumed to develop a large expectation value during inflation,
because of the large
negative Hubble-induced mass term.  After the inflation ends, 
the Hubble parameter gradually decreases. The AD field
starts coherent oscillations when $m_{x}$ eventually exceeds the Hubble
parameter.  At this time $H=H_{\rm osc}\simeq m_{x}$, a huge baryon asymmetry
is generated by the B-violating operators in the scalar potential
(\ref{scalar-potential}).

One can easily get the ratio of baryon to $\phi$-number density as
\begin{equation}
 \left(\frac{n_{B}}{n_{\phi}}\right)\simeq |a_{H}|\left(\frac{|\phi|_{0}}{M_{*}}\right)^2\delta_{\rm eff}\;,
\label{baryon-to-phi-ratio}
\end{equation}
where $\delta_{\rm eff}\equiv {\rm sin}({\rm arg}(a_{H})+4{\rm
arg}(\phi_{0}))$. This is fixed at $H=H_{\rm osc}$ and remains
constant until the $\phi$ fields decay into the gravitinos and SM
particles. Here, $\phi_{0}$ denotes the expectation value of the AD
field just after inflation.
We treat it as a free parameter for a
while, and assume that $|\phi|_{0}\lsim M_{*}$. Actually,
there exists a natural way to stop the AD field below the Planck
scale. This can be done by gauging the U(1)$_{B-L}$ symmetry.
Although the B-violating operators in Eq.~(\ref{kahler}) are invariants
of the U(1)$_{B-L}$
symmetry, the relevant flat direction field $\phi$ carries non-zero $B-L$
charges and can be lifted by the $D$-term potential.
In this case, the initial amplitude of the $\phi$ field is given by the $B-L$
breaking scale, $|\phi|_{0}\simeq v_{B-L}$~\cite{FHYB-L}.

In the present scenario, the quadratic potential supplied by the SUSY-invariant
mass term of the extra matter field $\bar{D}_{x}$ forbids the formation
of {\it stable} Q-balls. However, it cannot forbid the formation of {\it
unstable} ones. This is because radiative corrections
of SU(3)$_{C}$ and U(1)$_{Y}$ gauge interactions give rise to a
negative contribution to the
mass term $m_{x}^2 |\phi|^2$. At the one-loop level, the potential is
given by
\begin{equation}
 V= m_{x}^2\left(1+K \;{\rm log}\left(\frac{|\phi|^2}{M_{G}^2}\right) \right)|\phi|^2
\;,
\label{radiative-correction}
\end{equation}
where
\begin{equation}
 K\simeq -\frac{1}{4\pi}\left(
\frac{16}{3}\alpha_{s}+\frac{4}{15}\alpha_{1}
\right)\;.
\label{K-factor}
\end{equation}
Here, $M_{G}$ denotes a renormalization scale at which $m_{x}$ is
defined. In this expression, we neglect the corrections from the
possible Yukawa interactions, which should be
small enough to avoid large FCNC interactions.
In the following discussion, we use $K=-0.03$ for a representative
value.~\footnote{It depends on the matter contents of the model
and the scale of $|\phi|$. However, as we will see, the resultant baryon
asymmetry
and the mass density of dark matter depend only weekly on the precise
value of the  $K$-factor.}
These radiative
corrections make the potential a little bit flatter than the quadratic one,
which causes spatial instabilities and the formation of
unstable Q-balls inevitable.
A detailed discussion of the properties of the unstable Q-ball
can be found in Ref.~\cite{FH}.

For the purpose of calculating  the resultant baryon asymmetry,
we need the decay temperature of the Q-ball.
In the present model, the resultant Q-ball charge, $Q$, is estimated as
\begin{equation}
 Q=\bar{\beta}\left(\frac{|\phi|_{0}}{m_{x}}\right)^2\epsilon_{c}\;,
\label{q-ball-charge}
\end{equation}
where $\beta\simeq 6\times 10^{-3}$ and $\epsilon_{c}\simeq 0.01$ are
determined by the detailed lattice simulations~\cite{KK}, which are
independent of $|\phi|_{0}$.  The size of the charge that evaporates
from the surface of a single Q-ball through interactions with thermal
backgrounds is about $\Delta Q\simeq 10^{18}$~\cite{thermal-e}.
Consequently, the produced Q-balls survive thermal evaporation as long
as $|\phi|_{0}\gsim 10^{14}\GEV$. The remaining charges are emitted
through its decay into light fermions. The decay rate was calculated
in Ref.~\cite{Q-evaporation} as
\begin{equation}
 \Gamma_{Q}\equiv -\frac{dQ}{dt}\lsim \frac{\omega^3{\cal{A}}}{192\pi^2}\;.
\label{decay-rate}
\end{equation}
Here, ${\cal{A}}=4\pi R_{Q}^2$ is the surface area of the Q-ball,
where $R_{Q}\simeq \sqrt{2}/(m_{x}\sqrt{|K|})$ is the Q-ball
radius; $\omega\simeq m_{x}$ is the effective mass
of the Q-ball per $\phi$-number~\cite{unstable-Q}.
In the present model, the above inequality is almost
saturated.  The detailed discussion of the decay
process and the condition for this saturation will be
found in the next section.

From Eq.~(\ref{decay-rate}), the Q-ball decay temperature can be derived
as follows:
\begin{eqnarray}
 T_{d}&=&\frac{\eta}{\sqrt{48|K|\pi}}\left(\frac{90}{\pi^{2}g_{*}(T_{d})}\right)^{1/4}
\left(\frac{m_{x}M_{*}}{Q}\right)^{1/2}\nonumber\\
&\simeq&2\GEV\times\eta\left(\frac{0.03}{|K|}\right)^{1/2}\left(\frac{m_{x}}{1\TEV}\right)^{1/2}
\left(\frac{10^{20}}{Q}\right)^{1/2}\;,
\label{decay-temp1}
\end{eqnarray}
where $\eta\sim 1$ denotes the ambiguity coming from an inequality of
the decay rate in Eq.~(\ref{decay-rate}); $g_{*}(T_{d})$ is the
total number of  effectively massless
degrees of freedom at $T=T_{d}$.
From Eq.~(\ref{q-ball-charge}), we can also write $T_{d}$ in terms of the
initial amplitude of the $\phi$ field:
\begin{equation}
 \frac{T_{d}}{m_{x}}=\frac{\eta}{\sqrt{48\pi|K|\bar{\beta}\epsilon_{c}}}\left(
\frac{90}{\pi^2 g_{*}(T_{d})}\right)^{1/4}\left(\frac{m_{x}M_{*}}{|\phi|_{0}^{2}}\right)^{1/2}\;.
\label{decay-temp2}
\end{equation}

The low decay temperature indicated in Eq.~(\ref{decay-temp1}) makes it
much easier for the Q-balls to dominate the energy density of the
Universe. The condition  for the Q-ball dominance is given by
\begin{equation}
 T_{R}>3 T_{d}\left(\frac{M_{*}}{|\phi|_{0}}\right)^2\;.
\label{dominance-condition}
\end{equation}
In the remainder of this paper, we consider the case where the above
condition is satisfied. The corresponding region in the
$(|\phi|_{0}{\mbox{--}}T_{R})$ plane with thermal effects taken into
account is presented in Ref.~\cite{FY}. Basically, there is a wide
parameter space for $|\phi|_{0}\gsim 10^{15}\GEV$. In the case of
$|\phi|_{0}\simeq M_{\rm GUT}\;(\simeq 2\times 10^{16}\GEV)$, for
example, $10^2\GEV\lsim T_{R}\lsim 10^{10}\GEV$ is the viable range of
the reheating temperature, which satisfies 
Eq.~(\ref{dominance-condition}). Here, the upper bound on the
reheating temperature comes from the condition to avoid the early
oscillation of the AD field by the thermal effects~\cite{FY}.

In the case of the Q-ball dominance, the resultant baryon asymmetry is
given by the following simple formula:
\begin{equation}
 \frac{n_{B}}{s}=\frac{\rho_{Q}}{s}\left(\frac{n_{\phi}}{\rho_{Q}}\right)
\left(\frac{n_{B}}{n_{\phi}}\right)=\frac{3}{4}\frac{T_{d}}{m_{x}}|a_{H}|\left(
\frac{|\phi|}{M_{*}}\right)^2\delta_{\rm eff}\;,
\label{B-asymmetry1}
\end{equation}
where $s$, $\rho_{Q}$ are the entropy density and the energy density
of the Q-ball in the Universe, respectively.
In terms of the density parameter of the baryon number, 
it is given by 
\begin{equation}
\Omega_{B}h^2\simeq 0.02\times \eta \left(\frac{0.03}{|K|}\right)^{1/2}
\left(\frac{m_{x}}{1\TEV}\right)^{1/2}
\left(\frac{|\phi|_{0}}{10^{16}\GEV}\right)\left(\frac{|a_{H}|\delta_{\rm eff}}{0.02}\right)\;,
\label{B-density-parameter}
\end{equation}
where $h$ is the present Hubble parameter in units of $100\;{\rm
km}\;{\rm sec}^{-1}\;{\rm Mpc}^{-1}$, and
$\Omega_{B}\equiv\rho_{B}/\rho_{c}$;
$\rho_{B}$ and $\rho_{c}$ are the energy density of the baryon and the
critical density in the present Universe. 
Here, we have expressed the decay temperature $T_{d}$ of the 
Q-ball with the initial amplitude of the AD field and $m_{x}$ by using 
Eq.~(\ref{decay-temp2}). 
The corresponding decay
temperature of the Q-ball can be estimated as $T_{d}\approx 200\MEV$ 
for $|\phi|_{0}\approx 10^{16}\GEV$.
Note that the resultant baryon asymmetry is totally independent of the
reheating temperature of inflation and the detailed history of the
Universe~\cite{FH}, since the final entropy density is solely produced via the
late-time decays of Q-balls.
The observed baryon asymmetry, $\Omega_{B}h^2\simeq 0.02$, suggests  the initial amplitude of the
AD field to be  $10^{15}\GEV\lsim |\phi|_{0}\lsim 10^{17}\GEV$, which surprisingly coincides with
the $B-L$ breaking scale
suggested from the see-saw mechanism~\cite{see-saw}.~\footnote{
The mass of the right-handed neutrino $M_{R}$ is given by
$M_{R}=\lambda v_{B-L}$, where $\lambda$ is some coupling constant.
This fact and the atmospheric-neutrino data 
naturally suggest $v_{B-L}\sim M_{\rm GUT}$.} 
\section{Gravitino Abundance}
So far, we have concentrated on the calculation of the resultant baryon
asymmetry in a Q-ball-dominated Universe. Now, let us turn our
attention to the calculation of the relic gravitino abundance.
In the present scenario,
there are two stages in the gravitino production. The first one is the
well known thermal gravitino production, which takes place at the
reheating epoch of inflation.
The second one is the non-thermal gravitino production from the late-time
decays of the Q-balls.

\subsection{Thermal Production}

First, let us estimate the abundance of thermal relics of gravitinos.
It was calculated in detail in Refs.~\cite{G-problem} and is given by
\begin{equation}
 \frac{n_{3/2}}{s}={\rm Min}\left\{\frac{45}{2\pi^{2}g_{*}(T_{C})}\frac{\zeta(3)}{\pi^{2}}
\left(\frac{3}{2}\right)\;\;,
\;\;\frac{1}{4}\left(\frac{90}{\pi^{2}g_{*}(T_{R})}\right)^{3/2}
\left(\frac{\zeta(3)}{\pi^{2}}\right)^{2}T_{R}M_{*}\vev{\Sigma_{\rm scatt}v_{\rm rel}}
\right\}\;,
\label{T-gravitino-initial}
\end{equation}
where
\begin{equation}
\vev{\Sigma_{\rm scatt}v_{\rm rel}}\approx 5.9\;\frac{g_{3}^{2}m_{\widetilde{G}}^2}{m_{3/2}^{2}M_{*}^{2}}
\label{cross-section}
\end{equation}
is the thermally averaged gravitino production cross section. Here,
$m_{\widetilde{G}}$ denotes the mass of the gluino. The first quantity in
Eq.~(\ref{T-gravitino-initial}) denotes the gravitino abundance when it
is thermalized; $T_{C}$ is the ``decoupling'' temperature of the gravitino,
which is roughly given by
\begin{equation}
 \frac{\zeta(3)}{\pi^2}\;\frac{T_{C}^{3}\vev{\Sigma_{\rm scatt}v_{\rm rel}}}{H(T_{C})}=\frac{3}{2}
\;.
\label{decoupling}
\end{equation}
If the reheating temperature is higher than the mass of messenger
(s)quarks $M_{m}$, gravitinos are easily thermalized. In this case,
$T_{C}$  should be replaced by the smaller one between $M_{m}$ and the
solution $T_{C}$ of Eq.~(\ref{decoupling}).

Thermally produced gravitinos presented in
Eq.~(\ref{T-gravitino-initial}) are significantly diluted by the
subsequent entropy production associated with the Q-ball decays.
The dilution factor via the late-time decays of Q-balls is given by~\cite{FH}
\begin{equation}
 \frac{1}{\Delta}=3\frac{T_{d}}{T_{\rm ini}}\left(
\frac{M_{*}}{|\phi|_{0}}
\right)^2\;,
\label{dilution-factor}
\end{equation}
where
\begin{equation}
 T_{\rm ini}={\rm Min}\left[
T_{R}\;\;,\;\;\sqrt{m_{x}M_{*}}\left(
\frac{90}{\pi^{2}g_{*}}
\right)^{1/4}
\right]\;.
\label{Tini}
\end{equation}
The second term in the square bracket of  Eq.~(\ref{Tini}) denotes the case where the AD field
starts coherent oscillation after the reheating process of inflation.
In this case, it is difficult for the AD field with
$|\phi|_{0}\simeq M_{GUT}$ to escape strong thermal
effects, which modify the resultant baryon asymmetry and  the
properties of the produced Q-balls. Therefore, we concentrate on the
other case, and assume that
\begin{equation}
 T_{R}\lsim \sqrt{m_{x}M_{*}}\left(\frac{90}{\pi^2 g_{*}}\right)^{1/4}
\approx 2\times 10^{10}\GEV
\end{equation}
in the remainder of this paper.

First, let us consider the case where the gravitinos are thermalized.
From Eqs.~(\ref{T-gravitino-initial}) and (\ref{dilution-factor}), we obtain
\begin{equation}
 \frac{n_{3/2}}{s}=1.25 \times \frac{T_{d}}{g_{*}(T_{C})T_{R}}\left(
\frac{M_{*}}{|\phi|_{0}}\right)^{2}\;.
\end{equation}
In terms of the density parameter, it becomes
\begin{equation}
 \Omega_{3/2}h^2\simeq 0.2\times\eta \left(\frac{0.03}{|K|}\right)^{1/2}
\left(\frac{300}{g_{*}(T_{C})}\right)\left(\frac{m_{3/2}}{1\MEV}\right)\left(
\frac{10^{8}\GEV}{T_{R}}\right)\left(\frac{m_{x}}{1\TEV}\right)^{3/2}
\left(\frac{10^{16}\GEV}{|\phi|_{0}}\right)^{3}\;.
\label{density-Tcase}
\end{equation}
Here, we have used the relation in Eq.~(\ref{decay-temp2}).  The
required mass density of CDM, $\Omega_{3/2}h^2\simeq 0.1\mbox{--}0.2$, 
can be easily explained by
adjusting the reheating temperature of inflation. Note that this
adjustment does not change the resultant
baryon asymmetry, since it is independent of the reheating temperature
(see Eq.~(\ref{B-density-parameter})).

A much more interesting situation arises if the gravitinos are not
thermalized in the early Universe. If this is the case, 
from Eqs.~(\ref{T-gravitino-initial}) and (\ref{dilution-factor}), 
the resultant gravitino abundance is given by
\begin{equation}
 \frac{n_{3/2}}{s}=0.31\times \frac{T_{d}M_{*}}{g_{*}(T_{R})^{3/2}
\vev{\Sigma_{\rm scatt}v_{\rm rel}}^{-1}}\left(\frac{M_{*}}{|\phi|_{0}}\right)^2\;.
\label{gravitino-nonTcase}
\end{equation}
One can see that the resultant gravitino abundance is also independent
of the reheating temperature of inflation.
In terms of the density parameter, this is written as
\begin{equation}
 \Omega_{3/2}h^2\simeq 0.16\times \eta \left(\frac{250}{g_{*}(T_{R})}\right)^{3/2}
\left(\frac{0.03}{|K|}\right)^{1/2}\left(\frac{1\MEV}{m_{3/2}}\right)\left(
\frac{m_{\widetilde{G}}}{1\TEV}\right)^2\left(\frac{m_{x}}{1\TEV}\right)^{3/2}
\left(\frac{M_{\rm GUT}}{|\phi|_{0}}\right)^3\;.
\label{density-nonTcase}
\end{equation}
Therefore, in this case, both the baryon asymmetry and the mass
density of gravitino dark matter are solely determined by
electroweak-scale parameters and by the initial amplitude of the AD
field. Encouragingly, both quantities can be consistent with
observations in a reasonable parameter space.~\footnote{Contributions
  to the mass density of dark matter from the non-thermal gravitino
  production should be subdominant. See the next subsection.}

In fact, we can derive a novel relation between the mass density of
baryons and that of the gravitino dark matter in terms of low-energy parameters:
\begin{eqnarray}
&& (\Omega_{B}h^2)\times (\Omega_{3/2}h^2)^{\frac{1}{3}}\nonumber\\
&&\;\; \simeq 1.1\times 10^{-2}\;\eta^{\frac{4}{3}}\left(\frac{250}{g_{*}(T_{R})}\right)^{\frac{1}{2}}\left(\frac{1\MEV}{m_{3/2}}\right)^{\frac{1}{3}}
\left(\frac{m_{\widetilde{G}}}{1\TEV}\right)^{\frac{2}{3}}
\left(\frac{m_{x}}{1\TEV}\right)\left(\frac{0.03}{|K|}\right)^{\frac{2}{3}}
\left(\frac{|a_{H}|\delta_{\rm eff}}{0.01}\right)\;,\nonumber\\
\label{B-D-ratio}
\end{eqnarray}
which is quite consistent with the observation.
We should stress that the above combination of the 
baryon asymmetry and the gravitino dark-matter density is
independent of the unknown high energy physics parameters, such 
as $|\phi|_{0}$ and $T_{R}$. 
There remain only a weak dependence on the effective degrees of freedom,
$g_{*}(T_{R})$, and a natural assumption on the coupling 
constant and the effective CP-violating phase,
$|a_{H}|\delta_{\rm eff}={\cal{O}}(0.01\mbox{--}0.1)$.

We can obtain another interesting relation by taking a ratio of 
the mass density of baryon to that of dark matter:
\begin{equation}
\frac{\Omega_{B}}{\Omega_{3/2}}
\simeq 0.2\times \left(\frac{m_{3/2}}{1\MEV}\right)
\left(\frac{1\TEV}{m_{x}}\right)
\left(\frac{1\TEV}{m_{\widetilde{G}}}\right)^{2}
\left(\frac{g_{*}(T_{R})}{250}\right)^{3/2}
\left(\frac{|\phi|_{0}}{M_{\rm GUT}}\right)^{4}
\left(\frac{|a_{H}|\delta_{\rm eff}}{0.02}\right)\;.
\end{equation}
As one can see, 
if the $B-L$-breaking scale (and hence $|\phi|_{0}$) 
is around the GUT scale, the present 
scenario predicts a successful relation between baryon asymmetry 
and dark matter.

\begin{figure}[ht!]
 \centerline{\psfig{figure=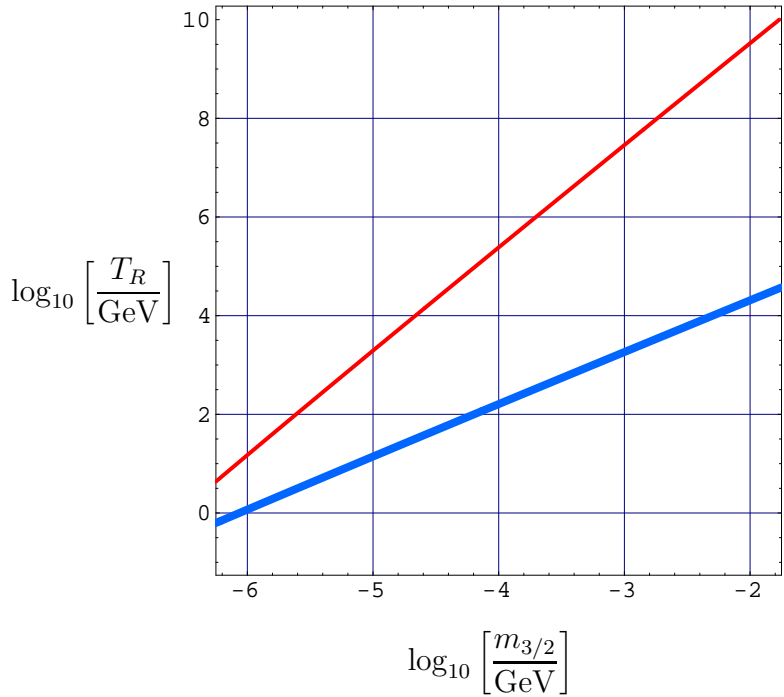,height=8cm}}
 \begin{picture}(0,0)
  \put(50,130){${\rm log}_{10}\left[\displaystyle{\frac{T_{R}}{{\rm GeV}}}\right]$}
  \put(200,-10){${\rm log}_{10}\left[\displaystyle{\frac{m_{3/2}}{{\rm GeV}}}\right]$ }
 \end{picture}
\vspace{0.8cm}
 \caption{The parameter space where the mass density of gravitino 
CDM is determined independently of the reheating temperature of inflation.
 Below the blue (thick) line, the mass density of gravitino is too
 small to explain the required mass density, $\Omega_{3/2}h^2<0.05$.
Above the red (thin) line, the gravitinos are thermalized by scattering
 interactions. The region between the two lines, the required mass density of
 gravitino dark matter, can be explained  by
 Eq.~(\ref{density-nonTcase}).
Here, we have assumed that the reheating temperature $T_{R}$ is smaller
 than
the messenger (s)quark masses.}
 \label{FIG-T-gravitino}
\end{figure}

In Fig.~\ref{FIG-T-gravitino}, we show a parameter space where we can
explain the gravitino CDM independently of the reheating temperature
of inflation. (As we have mentioned before, if the reheating temperature
is higher than the mass of the messenger (s)quarks, the gravitino
abundance should be calculated by Eq.~(\ref{density-Tcase}).)

Finally, we add a brief comment on the possible isocurvature
perturbations. In the present scenario,
the origin of the CDM is the inflaton decay, but the dominant entropy is
produced by the late-time decays of Q-balls.
Thus, the density fluctuation of the CDM contains partially 
isocurvature perturbations if the dominant-density 
perturbations are supplied by the fluctuation
of the AD field during inflation~\cite{Moroi-gumi}.
In order to avoid any
disagreement with observations, the Hubble parameter during
inflation should satisfy $H_{I}\lsim 10^{-5}|\phi|_{0}\approx
10^{10\mbox{--}12}\GEV$.~\footnote{
In this case, the dominant-density perturbations are
generated by the inflaton, and there remain small
isocurvature perturbations in the baryonic sector.} 
However, this is not a strong constraint, and
it can be easily satisfied, for example,
in hybrid and new inflation models~\cite{inflation-model}.
It will be an interesting indication of the present model,
if a small deviation from the purely adiabatic density perturbations
is indeed confirmed in the future experiments, such as MAP~\cite{MAP} and 
PLANCK~\cite{PLANCK}.

\subsection{Non-Thermal Production}
In this subsection, we discuss the non-thermal gravitino production
from the late-time decays of Q-balls.
We first estimate the resultant non-thermal gravitino abundance,
and then discuss experimental constraints and phenomenological
implications.

As we have mentioned in Section~\ref{B-asymmetry}, the AD field stored
inside the Q-ball decays into its superpartner and a gaugino, or
a quark/lepton and a higgsino,
through gauge or Yukawa interactions, respectively.
The following condition is necessary for the AD field to decay into
light fermions with almost saturated rate $(\eta\sim 1)$ in Eq.~(\ref{decay-rate}):
\begin{equation}
 f|\phi|_{c}\gsim \omega\;(\simeq m_{x})\;,
\label{decay-condtion}
\end{equation}
where $f$ denotes the corresponding gauge or Yukawa coupling constant,
and $|\phi|_{c}$ is the amplitude of the AD field at the centre of the
Q-ball. We can easily calculate the typical amplitude of the AD field
inside the Q-ball from its charge and radius, which turns out to be
$|\phi|_{c}\approx 10^{12}\GEV$ for $|\phi|_{0}\simeq M_{\rm GUT}$ and
clearly satisfies the above condition.

This is not the case for the decay of the AD field into the gravitino.
The relevant interaction is given by
\begin{equation}
 {\cal L}\supset \frac{m_{\chi_{\phi}}^2-m_{\phi}^2}{\sqrt{3}m_{3/2}M_{*}}(\barr{\psi}
\chi_{\phi})\phi^{*}+\rm{h.c.}\;,
\label{gravitino-interaction}
\end{equation}
where $\chi_{\phi}$ denotes the superpartner of the AD field $\phi$, and
$\psi$ is the longitudinal component of the gravitino ($\sim$ the Goldstino).
In the region where $|\phi|\gsim M_{m}$, the mass difference of the AD
field and its superpartner is given by~\cite{GMM} 
\begin{equation}
 m_{\phi}^2-m_{\chi_{\phi}}^2\simeq m_{\rm soft}^2\frac{M_{m}^2}{|\phi|^2}\;,
\label{mass-difference}
\end{equation}
where $m_{\rm soft}\simeq 1\TEV$ is the soft
SUSY-breaking mass of the AD field below the messenger scale $M_{m}$.
As a result, the branching ratio of the AD field that directly decays into the
gravitino is suppressed by the following factor with respect the
decay rate given in Eq.~(\ref{decay-rate})~\cite{Q-evaporation}:
\begin{equation}
 \frac{m_{\rm soft}^2 M_{m}^2/|\phi|}{\sqrt{3}m_{3/2}M_{*}}\times\frac{1}{\omega}\;.
\label{suppression}
\end{equation}
Below the messenger scale $|\phi|\lsim M_{m}$,
the numerator in (\ref{suppression}) is just given by 
$m_{\rm soft}^{2}|\phi|$. Thus the branching ratio is 
maximized at $|\phi|\simeq M_{m}$.
The resultant gravitino
abundance directly generated from the Q-ball decay is given by
\begin{eqnarray}
 \Omega_{3/2}^{\rm direct}h^2&\approx&(\Omega_{B}h^2)\left(
\frac{n_{\phi}}{n_{B}}\right)\times \frac{m_{\rm soft}^2 M_{m}}{\sqrt{3}M_{*}m_{x}m_{p}}\nonumber\\
&\simeq&5\times 10^{-6}\left(\frac{(n_{\phi}/n_{B})}{10^6}\right)
\left(\frac{m_{\rm soft}}{1\TEV}\right)^2\left(\frac{1\TEV}{m_{x}}\right)
\left(\frac{M_{m}}{10^6\GEV}\right)\;,
\label{direct-gravitino-abundance}
\end{eqnarray}
where $m_{p}$ is the mass of a nucleon.
In order to explain the observed baryon asymmetry,
the ratio of $n_{\phi}$ to $n_{B}$ is given by
$(n_{\phi}/n_{B})\approx 10^6$
from  Eqs.~(\ref{baryon-to-phi-ratio}) and (\ref{B-density-parameter}).
Therefore, from the above equation, the mass density 
of gravitinos directly produced by the decays of the AD fields inside
the Q-balls is a negligible
contribution to the total mass density of dark matter, 
as long as $M_{m}\lsim 10^{10}\GEV$.

A  superparticle produced by the late-time decays of Q-balls
subsequently decays into the next-to-lightest supersymmetric particle 
(NLSP $\chi$), which is the bino or the lightest
stau in the minimal GMSB models.
The NLSP, in turn, decays into the gravitino and its superpartner with
the following decay width:
\begin{equation}
 \Gamma_{\chi}\simeq \frac{1}{48\pi}\frac{m_{\chi}^5}{m_{3/2}^2 M_{*}^2}\;,
\label{decay-width}
\end{equation}
where $m_{\chi}$ is the mass of the NLSP.
In order not to spoil the success of the Big Bang nucleosynthesis (BBN),
it must satisfy the following constraint~\cite{constraint-life-time}:
\begin{eqnarray}
 T_{3/2}&\equiv&\left(\frac{90}{\pi^2 g_{*}(T_{3/2})}\right)^{1/4}\sqrt{\Gamma_{\chi}M_{*}}\;
\nonumber\\
&\simeq& 5\MEV\left(\frac{m_{\chi}}{100\GEV}\right)^{5/2}
\left(\frac{1\MEV}{m_{3/2}}\right)\gsim 5\MEV\;.
\label{BBN-constraint}
\end{eqnarray}
This condition is satisfied for a gravitino with
$m_{3/2}\lsim\;(\mbox{a few})\MEV$ for a reasonable range of mass of $\chi$.

First, let us consider the case where
\begin{equation}
 H(T_{d})\;\left(=\sqrt{\frac{\pi^2 g_{*}(T_{d})}{90}}\frac{T_{d}^2}{M_{*}}
\right)\gg \Gamma_{\chi}\;.
\end{equation}
Here, the produced NLSPs have enough time to reduce their abundance
via annihilations into the following value~\cite{FH-org} before they decay into gravitinos:
\begin{equation}
 \frac{n_{\chi}}{s}=\sqrt{\frac{45}{8\pi^2 g_{*}(T_{d})}}\;\frac{\vev{\sigma v}_{\chi}^{-1}}{M_{*} T_{d}}\;,
\end{equation}
where $\vev{\sigma v}_{\chi}$ denotes the s-wave component of the
annihilation cross section of the NLSP.
As a result, the resultant gravitino abundance is given by
\begin{equation}
 \Omega_{3/2}^{\chi}h^2\simeq 2.8\times 10^{-2}\left(\frac{m_{3/2}}{1\MEV}\right)
\left(\frac{10^{-11}\GEV^{-2}}{\vev{\sigma v}_{\chi}}\right)
\left(\frac{100\MEV}{T_{d}}\right)\;.
\label{H-larger-gamma}
\end{equation}

\begin{figure}[ht!]
 \centerline{\psfig{figure=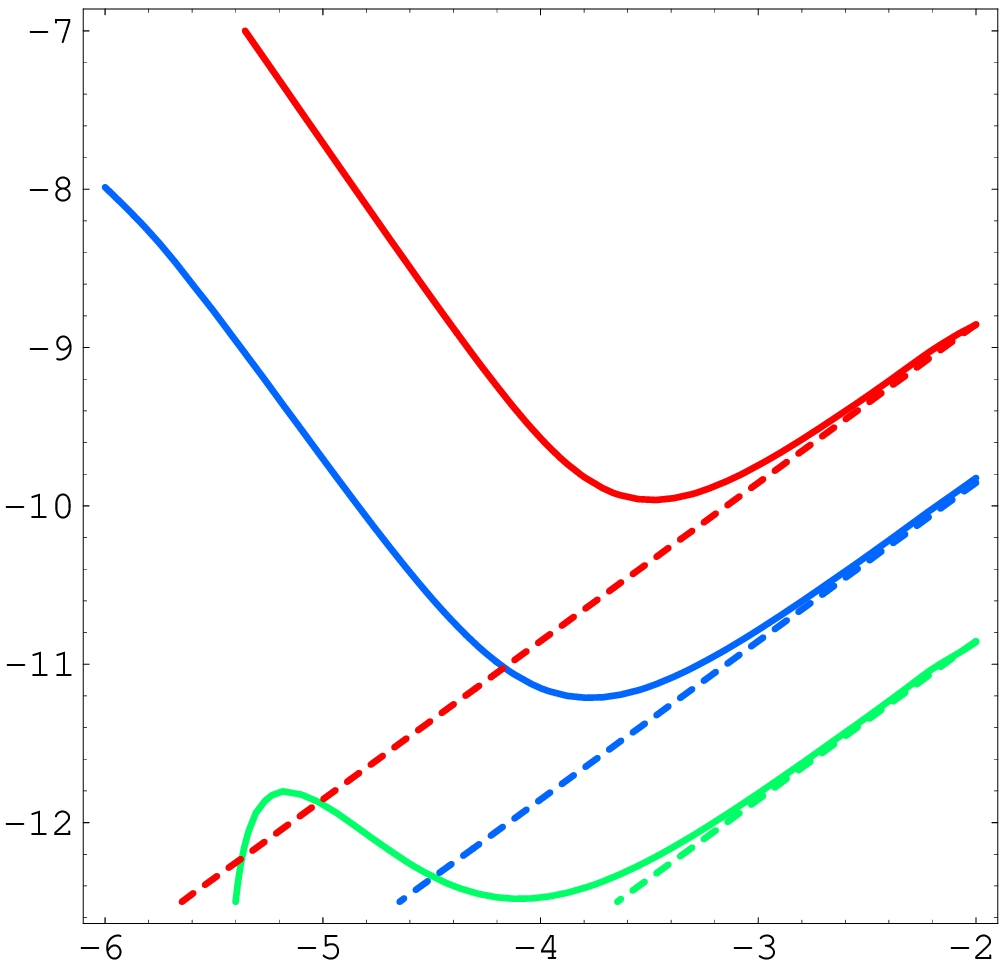,height=8cm}}
 \begin{picture}(0,0)
  \put(35,130){${\rm log}_{10}\left[\displaystyle{\frac{\vev{\sigma v}_{\chi}}{{\rm GeV^{-2}}}}\right]$}
  \put(200,-10){${\rm log}_{10}\left[\displaystyle{\frac{m_{3/2}}{{\rm GeV}}}\right]$ }
\small
\put(170,60){$0.15$}
\put(180,100){$0.015$}
\put(230,155){$0.0015$}
\normalsize
 \end{picture}
\vspace{0.8cm}
 \caption{The contour plot of the mass density of
non-thermal gravitinos $\Omega_{3/2}^{\chi}h^2$ in the
 $(m_{3/2}\mbox{--}\vev{\sigma v}_{\chi})$ plane.
The solid lines correspond to $\Omega_{3/2}^{\chi}h^2=0.15$,
 $0.015$ and $0.0015$ from the bottom
 up.
 The dashed lines represent the corresponding
 mass density of gravitinos calculated from Eq.~(\ref{H-larger-gamma}).}
 \label{FIG-non-thermal-production}
\end{figure}

In the case where $H(T_{d})\lsim \Gamma_{\chi}$, the estimation given in
Eq.~(\ref{H-larger-gamma}) is not valid, since the NLSPs produced by
the Q-ball decays do not have enough time to annihilate before
decaying into gravitinos. In this case, 
we have to solve the
following coupled Boltzmann equations~\cite{FH}:
\begin{eqnarray}
&& \dot{n}_{\chi}+3Hn_{\chi}=N_{\chi}\Gamma_{Q}n_{Q}^{\rm total}-\vev{\sigma v}_{\chi}n_{\chi}^2
-\Gamma_{\chi}n_{\chi}\;,\\
&&n_{Q}^{\rm total}+3 H n_{Q}^{\rm total}=0\qquad \;\qquad\qquad\qquad(\mbox{for}\;\;t\leq \tau_{d})\;,\\
&&n_{Q}^{\rm total}=0 \qquad\qquad\qquad\quad\qquad\qquad\qquad(\mbox{for}\;\;t\geq \tau_{d})\;,\\
&&H^2=\frac{1}{3 M_{*}^2}(\rho_{Q}+\rho_{\chi}+\rho_{\rm rad})\;,\\
&&\rho_{Q}=\epsilon_{c}^{-1}m_{\phi}(Q_{i}-\Gamma_{Q}t)
n_{Q}^{\rm total}\;,\\
&&\rho_{\chi}=m_{\chi}n_{\chi}\;,\\
&&\dot{\rho}_{\rm rad}+4 H\rho_{\rm rad}=\left(
\epsilon_{c}^{-1}m_{\phi}-N_{\chi}m_{\chi}
\right)\Gamma_{Q}n_{Q}^{\rm total}+m_{\chi}\vev{\sigma v}_{\chi}n_{\chi}^{2}+m_{\chi}
\Gamma_{\chi}n_{\chi}\;,\\
&&\dot{n}_{3/2}+3 H n_{3/2}=\Gamma_{\chi}n_{\chi}\;,
\label{bozmann}
\end{eqnarray}
where $n_{Q}^{\rm total}$ is the total number density of Q-balls,
which includes positively- and negatively-charged
Q-balls,~\footnote{As for the details, see Ref.~\cite{FH} and
  references therein.}  $m_{\phi}\simeq m_{x}$ is the mass of the AD
field, $n_{3/2}$ is the number density of gravitinos, $Q_{i}$ is the
initial charge of the Q-ball, $\tau_{d}\equiv Q_{i}/\Gamma_{Q}$ is the
lifetime of the Q-ball, and $N_{\chi}$ is the number of NLSPs produced
per baryon number, which is roughly equal to $\epsilon_{c}^{-1}$.
Note that the produced gravitinos behave as radiation for the era
relevant to the present calculation.  In
Fig.~\ref{FIG-non-thermal-production}, we show the contour plot of
$\Omega_{3/2}^{\chi}h^2$ by numerically solving the above Boltzmann
equations. Here, we have assumed that $T_{d}=100\MEV$, $m_{x}=1\TEV$,
$m_{\chi}=100\GEV$, $\epsilon_{c}^{-1}=N_{\chi}=100$.

Now, we discuss experimental constraints on the non-thermal
gravitino production.
First, let us  consider the constraint from the BBN. Even if the Q-balls
and the NLSPs decay into gravitinos well before the start of the BBN
(see Eq.~(\ref{BBN-constraint})),
the produced gravitinos are still ultra-relativistic during the era,
which may reduce too much the energy density during the formation of
light elements.

The momentum of the non-thermal gravitino redshifts with
temperature:~\footnote{In the case of the direct production of
  gravitinos from the AD fields stored in the Q-balls, $m_{\chi}$
  should be replaced by $m_{x}$.}
\begin{equation}
 p(T)\simeq\frac{m_{\chi}}{2}\left(
\frac{g_{*}(T)}{g_{*}(T_{g})}\right)^{1/3}
\frac{T}{T_{g}}\;,
\label{gravitino-monentum}
\end{equation}
where $T_{g}$ is the cosmic temperature at the time of 
the gravitino production.
Note that $T_{g}$ depends on the production mechanism and 
the relative size between $H(T_{d})$ and $\Gamma_{\chi}$, the decay rate
of the NLSP, and that $T_{g}$ is given by
\begin{eqnarray}
&& T_{g}=T_{3/2} \qquad\mbox{for}\quad (H(T_{d})\gsim\Gamma_{\chi})\nonumber\\
&& T_{g}=T_{d}   \qquad\;\;\; \mbox{for}\quad (H(T_{d})\lsim\Gamma_{\chi}\;
\;\mbox{and the direct gravitino production})\;.
\label{production-time}
\end{eqnarray}
Therefore, the contribution to the energy density of the 
radiation from the non-thermal gravitinos at 
temperature $T$ is estimated to
\begin{eqnarray}
&& \rho_{3/2}(T)\simeq\left(\frac{n_{3/2}}{s}\right)s(T)p(T)\nonumber\\
&&\;\;\simeq 7.8\times 10^{-3}\left(\frac{g_{*}(T)^{\frac{4}{3}}}{g_{*}(T_{g})^{\frac{1}{3}}}\right)
\left(\frac{m_{\chi}}{100\GEV}\right)\left(\frac{100\MEV}{T_{g}}\right)\left(
\frac{100\KEV}{m_{3/2}}\right)(\Omega_{3/2}^{\rm NT} h^2)T^4\;,
\label{radiation-component}
\end{eqnarray}
where $\Omega_{3/2}^{\rm NT}h^2$ denotes the present mass density of
non-thermal gravitino dark matter, which is $\Omega_{3/2}^{\rm direct}h^2$
or $\Omega_{3/2}^{\chi}h^2$.

In order not to affect the
expansion of the Universe during the BBN, the gravitino contribution to the
energy density of radiation should satisfy
\begin{equation}
 \frac{\rho_{3/2}}{\rho_{\nu}}\leq \delta N_{\nu}\;,
\label{delta-neutrino}
\end{equation}
where $\rho_{\nu}=(\pi^2/30)(7/4) T^4$ is the energy density of one
neutrino species.  Agreement with observations of light elements
requires $\delta N_{\nu}=0.2\mbox{--}1$~\cite{BBN}. This leads to the following
constraint:
\begin{eqnarray}
 \frac{\rho_{3/2}}{\rho_{\nu}}&\simeq& 1.4\times 10^{-3}\left(\frac{g_{*}(T)^{\frac{4}{3}}}{g_{*}(T_{g})^{\frac{1}{3}}}\right)\left(\frac{m_{\chi}}{100\GEV}\right)\left(\frac{100\MEV}{T_{g}}\right)
\left(\frac{100\KEV}{m_{3/2}}\right)\left(\frac{\Omega_{3/2}^{\rm NT}h^2}{0.1}\right)\nonumber\\
&\leq&0.2\mbox{--}1\;.
\end{eqnarray}
As can be seen, this constraint is very weak and almost always satisfied
for the  interested region of the parameter space.

Although the constraint coming from the BBN is very weak, the
non-thermal gravitinos should not make up a dominant portion of dark matter.
This is because they maintain relativistic velocity until just before the
matter-radiation equality. Actually, they are too hot to constitute
acceptable warm dark matter, even though they are 
a little bit cooler than neutrinos.
The current velocity of the non-thermal gravitino is given
by~\footnote{In the case of a direct production, $m_{\chi}$ should be
replaced by $m_{x}$.}
\begin{eqnarray}
 v_{0}&\simeq&\frac{m_{\chi}}{2 m_{3/2}}\left(\frac{g_{*}(T_{0})}{g_{*}(T_{g})}\right)^{1/3}
\frac{T_{0}}{T_{g}}\nonumber\\
&\simeq& 1.8\times 10^{-6}\left(\frac{1\MEV}{m_{3/2}}\right)
\left(\frac{m_{\chi}}{100\GEV}\right)\left(\frac{5\MEV}{T_{g}}\right)\;,
\label{current-velocity}
\end{eqnarray}
where $T_{0}=2.7\;{\rm K}$ is the current temperature of the cosmic
microwave background (CMB). Note that the velocity $v_{0}$ is 
independent of $m_{3/2}$
as long as $H(T_{d})\gsim \Gamma_{\chi}$,
since the gravitino mass dependence is
cancelled out by the $m_{3/2}$ dependence of its decay temperature $T_{3/2}$. 
This is the case when $m_{3/2}\gsim 100\KEV$ for the 
typical Q-ball decay temperature $T_{d}\approx 100\MEV$.
In the case of lighter
gravitino $m_{3/2}\lsim 100\KEV$, $v_{0}$ is inversely proportional to
the gravitino mass, since $T_{g}=T_{d}\approx 100\MEV$ is independent of
the gravitino mass (see Eq.~(\ref{production-time})).
On the other hand, $v_{0}\lsim 1.5\times
10^{-7}$ is required for warm dark matter to be consistent with
observations of the Lyman-$\alpha$ forest~\cite{lyman,non-thermal}, which corresponds to
the free-streaming scale $R_{f}\lsim 0.1\; {\rm Mpc}$.
In the present scenario, $R_{f}\approx 1\;{\rm Mpc}$ for $m_{3/2}\gsim
100\KEV$, and it even reaches $R_{f}\approx 100\;{\rm Mpc}$ for
$m_{3/2}\approx 1\KEV$.

Therefore, the contribution from the non-thermally produced
gravitinos should be a subdominant
component of dark matter; otherwise it would be in 
conflict with the observed large-scale structure of the Universe.
Although there has been derived no
accurate bound on the mass density of such ``hotter'' warm dark matter,
we can obtain a conservative bound by treating it as hot dark matter.
CMB measurements and galaxy cluster surveys already constrain  the mass
density of hot dark matter as $\Omega_{\rm hot}h^2\lsim 0.05$ at the
$95\%$ C.L.~\cite{wang-tegmark-zaldarriaga}.
From Eq.~(\ref{direct-gravitino-abundance}) and
Fig.~\ref{FIG-non-thermal-production}, one can see that
this constraint is satisfied as long as $M_{m}\lsim 10^{10}\GEV$ and
$\vev{\sigma v}_{\chi}\gsim \;(\mbox{a few})\times 10^{-12}\GEV^{-2}$.
In the case of the stau NLSP, the s-wave annihilation cross section is
of the order of $10^{-7}\GEV^{-2}~(100\GEV/m_{\chi})^2$, which clearly
satisfies the condition. Even in the case of the bino NLSP, $\vev{\sigma
v}_{\chi}\gsim 10^{-11}\GEV^{-2}$ is naturally obtained for relatively large
${\rm tan}\beta$ $(\gsim 15)$.
If the non-thermal gravitinos make up at least few per cent of the total
mass density of dark matter, data from the MAP or PLANCK satellites may
confirm its existence in the near future~\cite{hu-tegmark}.

\section{Conclusions and Discussion}
In this work, we have proposed a solution to the baryon asymmetry and
the dark
matter problems in the GMSB models.  We have  found that both problems
can be simultaneously solved if there exist extra matter multiplets
of a SUSY-invariant mass of the order of the $\mu$-term, which is 
suggested by several realistic SUSY GUT models utilizing some discrete
symmetries to realize the doublet--triplet splitting
in SUSY GUTs~\cite{SUSY-GUT}. The resultant baryon asymmetry is totally
independent of the reheating temperature of inflation, and the
required initial amplitude of the AD field is perfectly consistent
with the $B-L$ breaking scale $v_{B-L}\approx M_{\rm GUT}$
suggested from  neutrino-oscillation data.  
Furthermore, the abundance of thermal
gravitino dark matter is also independent of the reheating temperature
in a relatively wide parameter space. In that region, the abundance of
gravitino dark matter is solely determined by the electroweak-scale
parameters and the initial amplitude of the AD field.  Encouragingly, 
the required amplitude of the AD field is quite consistent
with the value required to explain the observed baryon asymmetry.

We have also discussed the non-thermal gravitino production via the
late-time decays of Q-balls. The constraints from the
BBN and the observation of the large-scale structures in the Universe can
be easily satisfied unless the messenger scale is too high, $M_{m}\gsim
10^{10}\GEV$, or the gravitino is relatively heavy, $m_{3/2}\gsim 10\MEV$.
In the case of the bino NLSP, the ``hotter'' warm dark 
matter supplied by
the non-thermal gravitino production is expected to be larger than a few
per cent of the total mass density of dark matter, whose existence may
be tested by MAP or PLANCK experiment in the near future~\cite{hu-tegmark}.

Finally, we add a comment on the effects of possible
non-renormalizable operators in the superpotential. The discrete
symmetry, which gives us a natural solution to the doublet--triplet
splitting problem, may allow non-renormalizable operators in the
superpotential, which lift the relevant flat direction. For example,
$Z_{4R}$ symmetry~\cite{MARU} allow non-renormalizable operators proportional to
$\phi^{6}$, such as $\propto (\bar{U}\bar{D}\bar{D})^2$. In the presence
of the gauged U(1)$_{B-L}$ symmetry, the operator is written as
$\delta W=(\lambda X/M_{*}^4)\phi^6$, where $\lambda$ is a coupling
constant, and $X$ is the Higgs field carrying $B-L$
charge $2$ to break the
$B-L$ symmetry spontaneously.~\footnote{
Assigning the $B-L$ charge $1$ to the $X$ field and adopting a
superpotential $\delta W=(\lambda X^2/M_{*}^5)\phi^6$ do not change
the main arguments in the following discussion.}
This operator lifts the $\phi$ field
at $|\phi|\approx M_{*}(m_{x}/\lambda \vev{X})^{1/4}$, which is of the same
order as the $B-L$-breaking scale. Hence, even if this operator
exists, it is still possible for the $\phi$ field to 
have its initial amplitude of the order of the $B-L$-breaking scale.
The contribution of this operator to the baryon
asymmetry is approximately given by $(n_{B}/n_{\phi})\simeq
(m_{3/2}/m_{x})\delta_{\rm eff}$, which is subdominant as long as
$m_{3/2}\lsim 1\MEV$ respect to that in Eq.~(\ref{baryon-to-phi-ratio})
(see also the discussion just below 
Eq.~(\ref{direct-gravitino-abundance})). 
Therefore, the results of the present model
are basically independent of the existence of such non-renormalizable
operators in the superpotential.

\section*{Acknowledgements}
M.F. thanks the Japan Society for the Promotion of Science for
financial support.  This work was partially supported by Grant-in-Aid
for Scientific Research (S) 14102004 (T.Y.).

\small


\end{document}